\documentclass[twocolumn,preprintnumbers,amsmath,amssymb,aps,pra,longbibliography]{revtex4-1}
\usepackage{amsmath,amssymb,graphicx,epstopdf,bm,soul,upgreek, natbib}
\usepackage{graphicx}
\usepackage{color}
\definecolor{OliveGreen}{rgb}{0,0.6,0}
\usepackage{dcolumn}
\usepackage{bm}
\usepackage{amssymb,float}
\usepackage{cancel}
\usepackage{ulem}
\usepackage{siunitx}
\usepackage[colorlinks,bookmarks=false,citecolor=blue,linkcolor=blue,urlcolor=blue]{hyperref}

\begin{document}
\title{Spin-polarized antichiral exciton-polariton edge states}

\author{Ruiqi Bao}
\author{S. Mandal}\email[Corresponding author:~]{subhaska001@e.ntu.edu.sg}
\author{Huawen Xu}
\author{Xingran Xu}
\author{R. Banerjee}
\author{Timothy C. H. Liew}\email[Corresponding author:~]{timothyliew@ntu.edu.sg}
\affiliation{Division of Physics and Applied Physics, School of Physical and Mathematical Sciences, Nanyang Technological University, Singapore 637371, Singapore}


\begin{abstract}
We consider theoretically a system of exciton-polariton micropillars arranged in a honeycomb lattice. The naturally present TE-TM splitting and an alternating Zeeman splitting, where the different sublattices experience opposite Zeeman splitting, shifts the Dirac points in energy, giving rise to  antichiral behavior. In a strip geometry having zigzag edges, two pairs of edge states exist and  propagate in the same direction (including the states at the opposite edges). The edge modes localized at the opposite edges have opposite spins (circular polarizations), which leads to co-propagating $``\pm"$ spin channels. The antichiral edge states are protected by  non-zero winding numbers and can propagate around a 60 degree bend without being reflected. We further compare the transport properties of these edge states with chiral edge modes and propose a scheme to realize them experimentally.
\end{abstract}

\maketitle

\section{Introduction} \label{sec:conclusions}
The quantum Hall effect \cite{PhysRevLett.45.494}, where free electron gas subjected to a perpendicular magnetic field shows quantized Hall conductance, has laid the platform for topological phases. A few years later, the Haldane model was proposed \cite{PhysRevLett.61.2015}, where the quantum Hall effect was shown without any net magnetic field. The Haldane model has gone onto become the backbone of many noble topological systems such as anomalous Chern insulators \cite{Jotzu2014,Zhao2020}, quantum spin-Hall insulators \cite{PhysRevLett.95.226801}, etc. 

Recently, a new type of topological phase, which shows antichiral edge states, has been introduced based on a modified-Haldane model \cite{PhysRevLett.120.086603}. In contrast to previous topological phases, systems with antichiral edge states do not host a bulk band gap. Instead, in a strip geometry both the edge states propagate in the same direction and counter-propagating states within the same energy window lie in the bulk. Although, the antichiral edge states reside along with the bulk modes, they were shown to be reasonably robust against disorder \cite{PhysRevLett.120.086603}. Because of their peculiar properties, the antichiral edge states have been under intense investigation in various system including in exciton-polaritonics \cite{PhysRevB.99.115423}, photonics \cite{PhysRevB.101.214102,PhysRevLett.125.263603,JianfengChen}, acoustics \cite{doi:10.1063/5.0050645,PhysRevLett.124.075501}, electric circuits \cite{Yang2021} and others \cite{PhysRevB.101.195133,Manna__2020,PhysRevResearch.2.012071,PhysRevB.103.195310,PhysRevB.104.L081401}. 

\begin{figure}
\centering
\includegraphics[width=0.95\columnwidth]{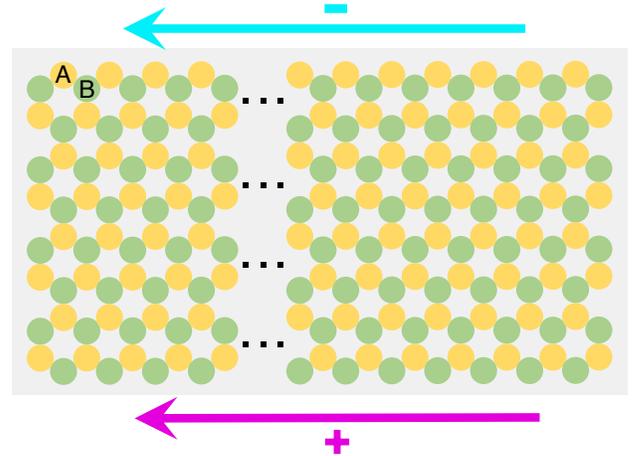}
\caption{Schematic of a honeycomb lattice consisting exciton-polariton micropillars. ``A" and ``B" correspond to the different sublattices of the system, which experience opposite Zeeman splittings. In presence of the sublattice dependent Zeeman splitting and the TE-TM splitting, spin-polarized antichiral edge states appear, where opposite edge attain opposite spins and propagate in the same direction.}
\label{Fig1}
\end{figure}

Although, the antichiral edge states have been explored over a variety of systems, the spin degree of freedom has not been associated with them till now. In this work, we propose theoretically spin-polarized antichiral edge states in an exciton-polariton system. Exciton-polaritons are quasi particles that arise due to the strong coupling between quantum well excitons and microcavity photons~\cite{RevModPhys.82.1489,RevModPhys.85.299,Byrnes2014}. Due to their excitonic component, polaritons exhibit strong nonlinearity, which has lead to a variety of optical devices \cite{Sanvitto2016,doi:10.1126/sciadv.abj6627}, optical computation \cite{PhysRevB.99.195301,PhysRevApplied.13.064074,Banerjee_2020,PhysRevApplied.17.024063}, and others \cite{PhysRevB.103.195302,Zvyagintseva2022}. The finite lifetime of the photons gives rise to interesting non-Hermitian physics. Circularly polarized photons provide a spin degree of freedom to the polaritons, which show many rich effects. For example, the naturally present energy splitting between the transverse electric (TE) and transverse magnetic (TM) modes (also known as the TE-TM splitting) acts as an effective magnetic field \cite{PhysRevLett.95.136601}, which leads to the realization of the optical spin-Hall effect both in the linear \cite{Leyder2007} and nonlinear \cite{PhysRevLett.109.036404} regimes. The excitonic component along with the spin degree of freedom leads to the Zeeman splitting in presence of perpendicular magnetic field \cite{PhysRevLett.105.256401}. These fascinating properties of polaritons make them ideal for studying topological phases in polariton lattices. 

Polariton Chern insulators \cite{PhysRevX.5.031001,PhysRevB.91.161413,PhysRevLett.114.116401,PhysRevB.98.075412,PhysRevB.97.081103,PhysRevApplied.12.064028} and related phases \cite{doi:10.1063/1.5018902,PhysRevApplied.12.054058} have been explored theoretically and later realized experimentally \cite{Klembt2018}. Strong nonlinear polariton-polariton interaction gives rise to different kinds of nonlinear topological behavior including topological solitons \cite{Kartashov:16,Gulevich2017,Pernet2022,PhysRevLett.118.023901}, interaction induced chiral edge states~\cite{PhysRevB.95.115415}, bistable topological insulators \cite{PhysRevLett.119.253904},  nonlinear higher-order topological insulators \cite{PhysRevLett.124.063901}, etc. In some cases, nonlinearity can alone induce topological phase transition \cite{PhysRevB.93.020502,PhysRevB.96.115453,PhysRevB.103.L201406}. The inherent non-Hermiticity of the polaritons has also lead to a variety of non-Hermitian topological behavior such as exceptional points \cite{Gao:2015uy,Gao:2018vv,Khurgin:20,PhysRevResearch.2.022051,doi:10.1126/sciadv.abj8905,PhysRevLett.127.107402,Krol:2022vg}, non-reciprocal transport \cite{PhysRevLett.125.123902,PhysRevB.104.195301}, skin effect \cite{doi:10.1021/acsphotonics.1c01425}, double-sided skin effect \cite{PhysRevB.103.235306}, non-Hermitian topological edge-modes \cite{Pickup2020,PhysRevResearch.2.022051}, etc. However, none of the above works can show topologically protected spin transport and the search for a system exhibiting robust spin transport is still an open question. 

Here, we consider exciton-polaritons in a honeycomb lattice in the presence of the TE-TM splitting and subjected to an alternating Zeeman splitting, where all the ``A" (``B") sublattices are considered to have positive (negative) Zeeman splitting. The same system with a uniform Zeeman splitting is known to give rise to a Chern insulating phase, where in a strip geometry, the bulk band gap hosts two pairs of counter-propagating edge states \cite{PhysRevLett.114.116401,PhysRevB.91.161413,Klembt2018}. Unlike the Chern insulator case, our present system does not show a bulk bandgap. Instead, the Dirac points split in energy. The non-trivial topology of the system connects the Dirac points with two pairs of zigzag edge modes with each pair residing on the opposite edges of a finite strip. Consequently, the edge modes including the opposite edges obtain the same sign of the group velocity. Unlike the Chern insulator case, here the system becomes antichiral, where all the edge modes propagate in the same direction. We further find that the edge states residing at the opposite edges of the strips have opposite circular polarization, which enables robust spin transport along the edges of the system. However, due to the fact that the edge modes reside with the counter-propagating bulk modes, the system is less robust compared to their Chern insulator counterpart. Our simulation shows that the system shows fair robustness against lattices defects and the efficiency of pulse propagation remains lower than but close to the Chern insulators.

\begin{figure*}[t]
\includegraphics[width=0.95\textwidth]{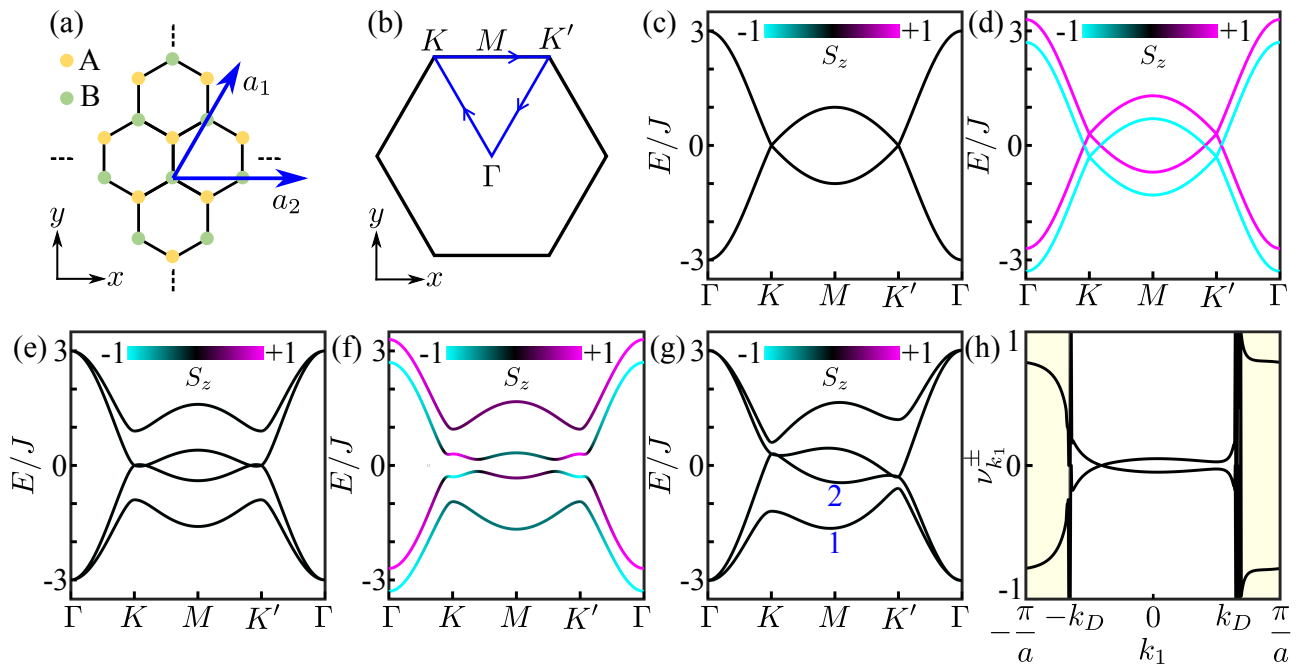}
\caption{(a)Honeycomb lattice consisting of two types of sublattices, ``A" and ``B". $a_1$ and $a_2$ are the primitive lattice vectors. (b) Lines connecting the high symmetry points of the 2D Brillouin zone, along which the band structures are calculated. (c) Band structure for a trivial Honeycomb lattice ($\Delta_T=\Delta_A=\Delta_B=0$). (d) Effect of Zeeman splitting on the honeycomb lattice band structure ($\Delta_T=0,~\Delta_A=\Delta_B=0.3J$). (e) Effect of the TE-TM splitting on the band structure ($\Delta_T=0.3J,~\Delta_A=\Delta_B=0$). (f) Chern insulating band structure with a non-trivial bulk bandgap ($\Delta_T=\Delta_A=\Delta_B=0.3J$). (g) Antichiral band structure  with the two valleys shifted in energy ($\Delta_T=\Delta_A=-\Delta_B=0.3J$). (h) Winding numbers as a function of $k_1$ for the antichiral bands, which become nontrivial for $k_1>|k_D|$, predicting the existence of two pairs of antichiral edge states in that momentum window.}
\label{Fig2}
\end{figure*}

\section{The model}
We start by considering two coupled micropillars ``A" and ``B" in presence of the Zeeman splitting and the TE-TM splitting. The corresponding Hamiltonian using the bispinor basis $\Psi=[\Psi_A^+,\Psi_A^-,\Psi_B^+,\Psi_B^-]^{T}$  can be expressed in the tight-binding limit as 
\begin{align}\label{Eq1}
\mathcal{H}_{AB}
=\begin{bmatrix}
            \Delta_A & 0  & J  & \Delta_T e^{-2i\theta}\\
            0   & -\Delta_A & \Delta_T e^{2i\theta} & J \\
            J & \Delta_T e^{-2i\theta} & \Delta_B & 0 \\
            \Delta_T e^{2i\theta} & J & 0 & -\Delta_B
\end{bmatrix}.
\end{align}
Here $J$ is the coupling strength between the pillars, $\Delta_T $ is the TE-TM splitting, $\theta$ is the orientation angle connecting the two pillars with respect to a reference axis, and $\Delta_{A(B)}$ is the Zeeman splitting corresponding to the pillar ``A"(``B"). We stress that while staggered potentials were considered in previous work predicting a valley-Hall effect~\cite{Bleu:2018vd}, here it is the Zeeman splitting that is staggered. We discuss explicit ways that it can be realized in section VI. The difference between staggered potentials and staggered Zeeman splitting is that in the case that a uniform magnetic field is applied together with a staggered potential, we would generally obtain the chiral edge states of a Chern insulator rather than antichiral ones. However, by properly tunning the parameters it is possible to completely gap out one of the valleys, leaving the other one gapped. Although, the system no longer remains a Chern insulator, it can act as a robust spin filter, where only one of the two spins are allowed to propagate along one of the edges. Such a regime is discussed in details in Ref.~\cite{PhysRevApplied.12.054058}.

Next, we arrange such pillar pairs in a honeycomb lattice structure and write down its corresponding Hamiltonian in the reciprocal space by taking periodic boundaries in both the directions. This can be achieved by writing $\Psi^{\pm}_{A,B,n} = \Psi^{\pm}_{A,B}e^{i\mathbf{k}.{\mathbf{x}}_n}$ where ${\mathbf{x}}_n$ is the position of each site and reducing the basis back to that of $[\Psi_A^+,\Psi_A^-,\Psi_B^+,\Psi_B^-]^{T}$.
\begin{align}\label{Eq2}
\mathcal{H}_{\mathbf{k}}
=\begin{bmatrix}
            \Delta_A & 0  & -g_{\mathbf{k}}J  & -g^+_{\mathbf{k}}\Delta_T \\
            0   & -\Delta_A &-g^-_{\mathbf{k}}\Delta_T & -g_{\mathbf{k}}J  \\
            -g^*_{\mathbf{k}}J &-{g^-_\mathbf{k}}^*\Delta_T& \Delta_B & 0 \\
            -{g^+_\mathbf{k}}^*\Delta_T & -g^*_{\mathbf{k}}J & 0 & -\Delta_B
\end{bmatrix},
\end{align}    
where $$g_{\mathbf{k}}=\sum_{n=1}^3\exp\left({-i\mathbf{k}.{\mathbf{r}}_n}\right)$$ and 
$$g^\pm_{\mathbf{k}}=\sum_{n=1}^3\exp\left({-i\left[\mathbf{k}.{\mathbf{r}}_n\mp 2\theta_n\right]}\right).$$
Here $\mathbf{r}_{n}$ represent the vectors connecting the three nearest ``B" sites from a single ``A" site (Fig.~\ref{Fig2}(a)) and $\theta_n=2\pi(n-1)/3$ are the angles of those vectors  with respect to one of them (say $\mathbf{r}_{1}$). 

To investigate the effect of $\Delta_T$, $\Delta_A$, and $\Delta_B$, we plot the band structure of the system along a line connecting the high symmetry points in the 2D Brillouin zone. Such a line is shown in Fig.~\ref{Fig2}(b). Before proceeding, we define the degree of circular polarization corresponding to a state as
\begin{align}\label{Eq3}
S_z=\frac{\left|\Psi^+_A\right|^2+\left|\Psi^+_B\right|^2-\left|\Psi^-_A\right|^2-\left|\Psi^-_B\right|^2}{\left|\Psi^+_A\right|^2+\left|\Psi^+_B\right|^2+\left|\Psi^-_A\right|^2+\left|\Psi^-_B\right|^2}.
\end{align}
The Hamiltonian of the system for  $\Delta_T=\Delta_A=\Delta_B=0$ corresponds to trivial Graphene. Consequently, we obtain the Graphene band structure by diagonalizing Eq.~(\ref{Eq2}), where the lower and upper band touch each other at the $K$ and $K^\prime$ valleys forming the Dirac cones (see Fig.~\ref{Fig2}(c)). In this case, the bands corresponding to the ``$\pm$" spins are degenerate or equivalently they are linearly polarized with $S_z=0$. This degeneracy can be lifted by making $\Delta_A=\Delta_B=\Delta$, which breaks the time-reversal symmetry (TR) and shifts the bands corresponding to the ``$\pm$" spins in energy by $2\Delta$ (see Fig.~\ref{Fig2}(d)). Next, we make $\Delta_T\neq0$ keeping $\Delta_A=\Delta_B=0$. In this case, the TR symmetry is restored and the bands become linearly polarized. One significant difference from the previous two cases is that the band structure is no longer linear near the $K$ and $K^\prime$ valleys (see Fig.~\ref{Fig2}(e)). Making $\Delta_T\neq0$ 
and $\Delta_A=\Delta_B\neq0$ gives rise to the non-trivial topological band structure with the system transiting to a Chern insulating phase having Chern number 2 \cite{PhysRevLett.114.116401,PhysRevB.91.161413,Klembt2018}. A nontrivial bulk band gap opens up due to the band inversion, which can be identified by the opposite values of $S_z$ of the second and third bands near the valleys (see Fig.~\ref{Fig2}(f)). Finally, we introduce the antichiral band structure in Fig.~\ref{Fig2}(g) by choosing $\Delta_T\neq0$ and $\Delta_A=-\Delta_B\neq0$. Unlike the Chern insulator band structure, the bulk bands are not gapped here. Instead the states near the $K$ and $K^\prime$ prime valleys shift in energy. The introduction of $\Delta_{A(B)}$ breaks the TR symmetry. However, unlike Figs.~\ref{Fig2}(d,f) the bulk bands do not have any preferred spins and stay linearly polarized. 

The antichiral system does not have a bulk band gap. Consequently, it is not possible to assign a Chern number to characterize the topology of the system. It is possible to calculate a relevant winding number. However, unlike all the previous antichiral systems, instead of one, two bands are involved here. Consequently, one needs to define the non-abelian Berry connection in order to calculate the winding number (see appendix). This is defined as \cite{PhysRevB.78.195424}
\begin{align}\label{Eq4}
\mathcal{F}^{m,n}_{k_1}=i\langle u^m_{k_1}(k_2)\left|\frac{\partial}{\partial k_2}\right|u^n_{k_1}(k_2)\rangle,
\end{align}
where $m,n=1,2$ corresponds to the lowest two bands as indicated in Fig.~\ref{Fig2}(g). $u^n_{k_1}(k_2)$ is the Bloch wave function. $k_{1}$ and $k_{2}$ are the wave vectors along the $a_1$ and $a_2$ directions, respectively (see Fig.~\ref{Fig2}(a)). The winding number is given as 
\begin{align}\label{Eq5}
\nu_{k_1}=\nu^+_{k_1}+\nu^-_{k_1}=\frac{1}{\pi}\oint_{\text{BZ}}dk_2 \text{Tr}[\mathcal{F}_{k_1}],
\end{align}
where $\nu_{k_1}$ is the total winding number of the system, $\nu^\pm_{k_1}$ correspond to the contribution from the two bands, and the integration is taken over the Brillouin zone along the ${k_2}$ direction for each value of $k_1$. In Fig.~\ref{Fig2}(h) the winding numbers are plotted as a function of $k_1$, which shows 
\begin{align}\label{Eq6}
\nu^+_{k_1}=-\nu^-_{k_1}\approx\left\{\begin{matrix}0,~\text{for}\left|k_1\right|<k_D\\
1,~\text{for}\left|k_1\right|>k_D\end{matrix}\right.
\end{align}
where $k_D$ is the position of the Dirac point. The total winding number of the system is always 0. However, $\nu^\pm_{k_1}$ deviates from 0 and becomes close to $\pm1$ as shown in the shaded region in Fig.~\ref{Fig2}(h). Non-zero winding corresponds to the non-trivial topology and predicts the existence of antichiral edge states \cite{PhysRevLett.120.086603}. Two non-trivial branches in the winding number predicts two pairs of antichiral edge modes. This is different from previously studied systems, where only a pair of antichiral edge modes are obtained. 

\begin{figure}[t]
\includegraphics[width=0.5\textwidth]{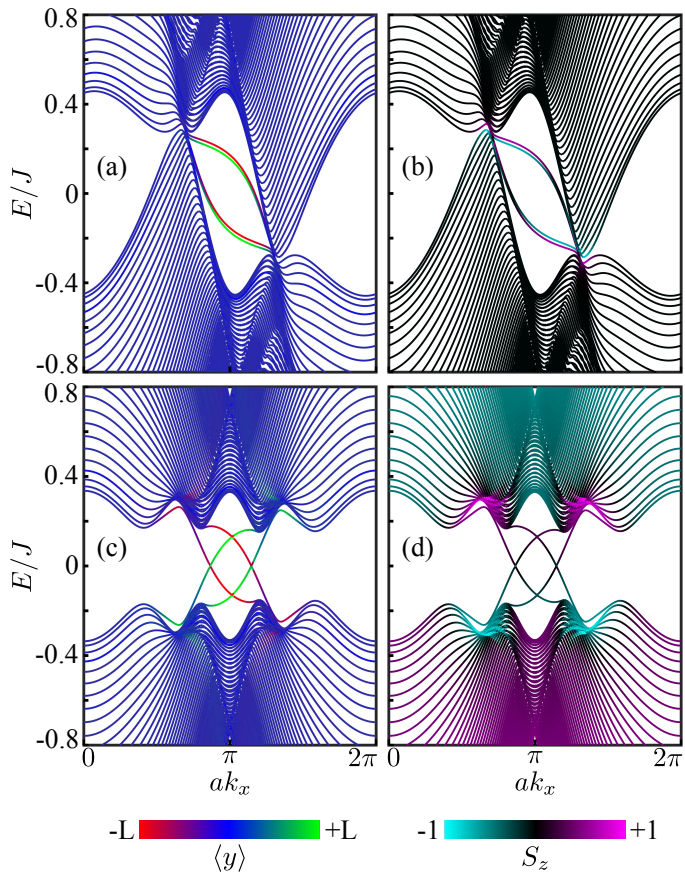}
\caption{Topological band structures for antichiral (a-b) and Chern insulator (c-d) cases. The states are color coded with respect to the contribution from the edges in (a,c) and with respect to the contribution from the spins in (b,d). Antichiral edge states are co-propagating having almost pure spins in a certain energy window. The chiral edge states in the Chern insulator case are counter-propagating and have mixed spins. Here $a$ is the periodicity along the $x$ direction, $\langle y\rangle$ represents of the mean position of the states along the $y$ axis with $-L$ and $+L$ being the boundaries. $S_z$ is defined in Eq.~(\ref{Eq3}). All the parameters are kept the same as those in Fig.~\ref{Fig2}.}
\label{Fig3}
\end{figure}    

\section{Topological band structure}
In this section, we provide  topological band structures  by considering the system periodic along the $x$ direction and truncated along the $y$ direction with the zigzag boundaries. In Figs.~\ref{Fig3}(a-b) the numerically calculated band structure for the antichiral system $(\Delta_T=\Delta_A=-\Delta_B)$ is shown. As predicted from the non-zero winding numbers, indeed two pairs of antichiral edge modes connecting the Dirac points appear and co-exist with the bulk modes. Generally, the edge modes localized at the opposite edges are degenerate. However, for better visualization, we have lifted the degeneracy of the edge modes by adding a inversion symmetry breaking term $m\sigma_z\otimes\sigma_o$ to the Hamiltonian in Eq.~(\ref{Eq2}). Here $\sigma_z$ is the $z$ component of the Pauli matrix, $\sigma_o$ is a $2\times2$ identity matrix, and $\otimes$ represents the tensor product. We have chosen a small $m=10^{-2}$ for Fig.~\ref{Fig3}(a-b), which does not alter any property of the system. This is necessary for  visualizing the properties of the edge modes from the band structure, otherwise the opposite edge modes in each pair would overlap  with each other making it difficult to distinguish them.

To study the properties of the edge modes, we color code the band structure as a contribution of localization in  Fig.~\ref{Fig3}(a) and as a contribution of $S_z$ in Fig.~\ref{Fig3}(b). All the edge modes have the same group velocity, which makes them propagate in the same direction. Such one directional propagation at the edges is balanced by the counter propagating bulk modes. Although, the bulk remains linearly polarized, the antichiral edge modes at the two opposite edges have opposite values of $S_z$ and in a particular energy window $S_z\rightarrow\pm1$. As we show in the next section, this leads to different spin channels in the same sample, where opposite spins propagate along the different edges of a finite sample in the same direction. 

In Figs.~\ref{Fig3}(c-d), we provide the Chern insulating band structure $(\Delta_T=\Delta_A=\Delta_B)$ for comparison. Unlike the antichiral system, here the bulk bands are gapped and two pairs of edge modes appear. Since the edge states at the different edges have opposite group velocities, they propagate in the opposite directions. Here the bulk bands are not linearly polarized and have some preferred value of $S_z$. Near the Dirac points, $S_z\rightarrow\pm1$. However, the edge modes deviate from the pure spins and acquire some mixed spins.

\section{Spin propagation}
\begin{figure}[t]
\centering
\includegraphics[width=0.47\textwidth]{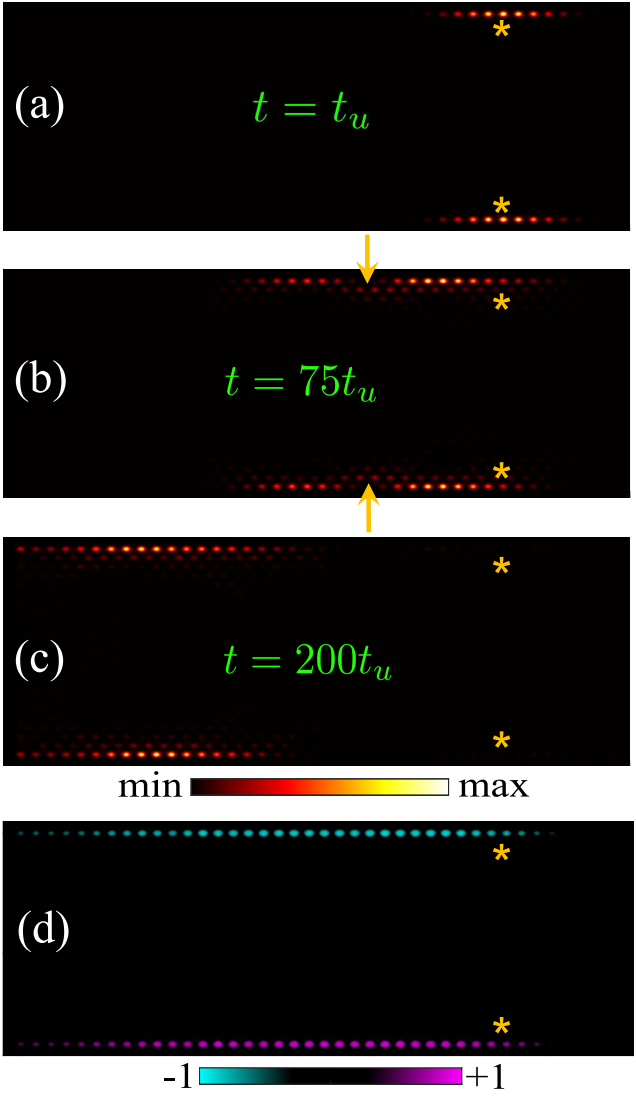}
\caption{(a-c) Polariton propagation under the effect of two pulses. The arrows in (b) represents the positions of two defects deliberately introduced by removing a site at each edge. Polaritons can propagate around the defect without significant backscattering. (d) Polariton propagation under  continuous coherent pumps. Polaritons with ``$-$" spin propagate along the upper edge, while the polaritons with ``$+$" spin propagate along the lower edge. The yellow stars correspond to the positions of linearly polarized coherent pulses (a-c) and pumps (d). Parameters: $F_0=J$, $\Gamma=0.002J$, $\sigma=3$, $\tau_0/\hbar=50/J$, $\hbar\omega_p = -0.207J$, $t_u = \hbar/J$ and $ak_p=3.42$. All other parameters are kept the same as those in Figs.~\ref{Fig3}(a-b).}
\label{Fig4}
\end{figure}

\begin{figure}[t]
\centering
\includegraphics[width=0.48\textwidth]{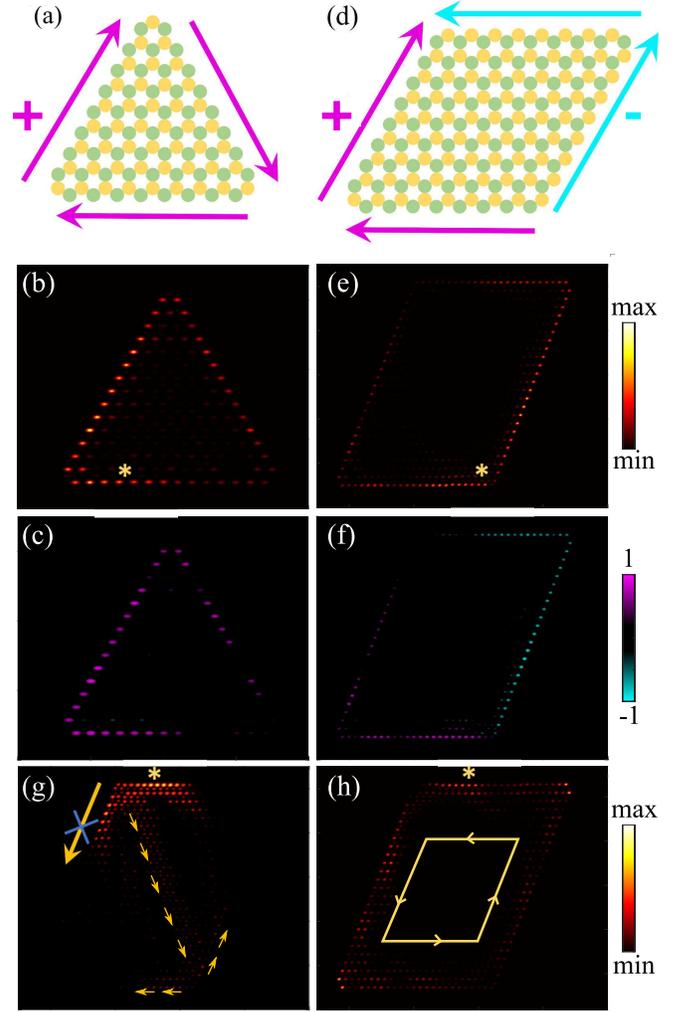}
\caption{Schematic diagram of antichiral system arranged in a triangle (a) and a parallelogram shape (d). 
The triangle shaped system has only one type of edge. (b-c) Only spin ``+" polaritons can propagate around the bending without being backscattered. (e-f) The  parallelogram consists of two types of zigzag edges and ``$\pm$" spins propagate along those edges, separately. (g) At the top left corner, where both edges meet, polaritons scatter into the bulk. (h) For the Chern insulator system, polaritons can propagate along the edges and do not scatter into the bulk. Each arm consists of 15 sites in (a-c) and 25 sites in (d-h). The yellow stars represent linearly polarized coherent pumps. Parameters: (a-g) All other parameters are kept the same as those in Fig.~\ref{Fig4}. (h) $\Delta_T=\Delta_A=\Delta_B=0.3J$, $\hbar\omega_p=0.07J$, and $ak_x=2.89$.} 
\label{Fig5}
\end{figure}

In this section, we investigate the polariton propagation in  the antichiral system and  find out their robustness by removing one site from each edge. We consider a stripe lattice formed by 12 unit cells along the $y$ direction and 20 unit cells along the $x$ direction. The upper (lower) edge of the system is ``A" (``B") type zigzag edge (see Fig.~\ref{Fig1}). The evolution of the polaritons under the coherent excitation is governed by the following time dependent Schr{\"o}dinger equation
\begin{equation}
i\hbar \frac{\partial\Psi_{\pm}}{\partial t} = H\Psi_{\pm} - i\Gamma\Psi_{\pm}+ F_\pm(x,y,t)\exp{\left[i\left(k_px-\omega_p t\right)\right]},
\end{equation}
where 
\begin{align}\label{PumpPulse}
F_\pm=F_0\left\{\begin{matrix}\exp\left[{-\frac{[(x-x_0)^2+(y-y_0)^2]}{2\sigma^2}-\frac{t^2}{2\tau_0^2}}\right],~\text{for pulse,}\\
\exp\left[{-\frac{[(x-x_0)^2+(y-y_0)^2]}{2\sigma^2}}\right],~\text{for pump}
\end{matrix}\right..
\end{align}
Here $H$ is the Hamiltonian representing the system which can be written by repeating the Hamiltonian $H_{AB}$, $F_{\pm}$ represents the circularly polarized components of Gaussian shaped coherent pulses or pumps having  amplitude $F_0$, energy $\hbar\omega_p$,  and momentum $k_p$; and $\Gamma$ is the decay rate of polaritons. The expression in Eq.~(\ref{PumpPulse}) represents the profiles of $F_\pm$ in space and in time.  $(x_0,y_0)$ represent the position of $F_\pm$ having widths $\sigma$ in space and $\tau_0$ in time. 

In Fig.~\ref{Fig4} the polariton propagation under two linearly polarized ($F_+=F_-$) pulses positioned at the two edges (represented by the yellow stars) are shown. As it can be seen from the band structure in Fig.~\ref{Fig3}(a-b) the edge modes have negative group velocity. Consequently, polaritons propagate from right to left along both the edges. To check the robustness, we have removed one site at each edge as indicated by the arrows in \ref{Fig4}(b). As it can be seen in Figs.~\ref{Fig4}(a-c) the polaritons can propagate around the defect without significant backscattering. However, the propagation around the defect is not perfect and in the next section, we have shown that the backscattering is around 10\%. Because of the presence of the decay term $\Gamma$, the intensity created by the pulse decreases exponentially. Consequently, for visulization purpose the intensity is renormalized at each time step in Figs.~\ref{Fig4}(a-c). Next, to show that the edges are circularly polarized, we use continuous pump instead of pulses. It should be noted that we use linearly polarized pumps to inject both $\pm$ polaritons. However, the system only allows $``+"$ polaritons to propagate along the lower edge and $``-"$ polaritons to propagate along the upper edge, creating two spin channels. 

The dynamics of the polaritons become more interesting if instead of a strip geometry, triangular and parallelogram shaped geometries are considered. In Fig.~\ref{Fig5}(a), (d) schematics of triangular and parallelogram shaped geometries are shown, respectively. For triangular geometry, one of the edges of the system can be considered as periodic  while the other edge does not appear in the structure. As a result, under a linearly polarized coherent pump, spin ``+" polaritons can propagate along the edges without being backscattered into the bulk (see Figs.~\ref{Fig5}(b-c)). This scenario changes for parallelogram shaped geometry as it consists of both type of edges. Consequently, $``\pm"$ polaritons propagate along different edges (see Figs.~\ref{Fig5}(e-f)). Due to the presence of $\Gamma$ the intensity of the polaritons decreases as they propagate away from the excitation spot. It should be noted that in both the structures, polaritons can go around the 60 degree bends perfectly. At 120 degree bends the two types of edge states meet (see the top left corner in Fig.~\ref{Fig5}(d)). Since both the edge states propagate in the same direction, polaritons can not continue propagating along the edges. Instead they scatter into the bulk. Such a situation is shown in Fig.~\ref{Fig5}(g), where the pump is positioned at the top left corner, where the opposite edges meet. For comparison, in Fig.~\ref{Fig5}(h) we have considered the case for the Chern insulator, which understandably does not scatter into the bulk and continues propagating along the edges.

\section{Efficiency comparison}
To estimate the backscattering while going around the defect, we consider the similar geometry as shown in Figs.~\ref{Fig4}(a-c), where the edge contains a defect. Next, we position a linearly polarized coherent pulse at the right end of the strip and  define the efficiency of the system as 
\begin{equation}
\eta = \frac{I_l-I_r}{I_l+I_r}
\end{equation}
where $I_l$ is the total intensity of the sites situated at the left of the defect (from both edge and bulk) and $I_r$ is the total intensity of sites situated at the right of the defect (from both edge and bulk). The intensities are considered at longer times when  the polaritons pass the defect. In Fig.~\ref{Fig6} the efficiency of the antichiral system is plotted in blue as a function of different system size $N$ as defined in the inset. For larger $N$, the efficiency stays near the 90\%, but never reaches 100\%, which is an indication that there is always around 10\% backscattering when the antichiral edge states go around a defect. This is understandable, as the antichiral edge modes reside together with the bulk modes in the same energy window, a defect can couple the edge and bulk modes. However, Since the edge modes are spatially separated from the bulk modes, the system still shows fair robustness and the efficiency remains around 90 \%. The efficiency decreases rapidly as $N$ decreases.

For comparison with the Chern insulator edge states, we also perform the same steps and obtain the intensity as shown in red. As expected, the Chiral edge modes of the Chern insulators show near perfect transmission around a defect and the efficiency remains around  100 \%. Similar to antichiral system the efficiency decreases rapidly for lower $N$. The low $\eta$ for lower $N$ is understandable from the fact that, being topological in nature, the robustness of the edge modes in both the systems is associated with the bulk properties. However, if sufficient bulk is not provided to the system, the system can not show the expected robustness. {We should notice that the nature of backscattering is different for the antichiral and Chern insulator case. For the antichiral system, the drop in efficiency arises due to the coupling of co-propagating antichiral edge states with the counter propagating bulk states. However, for the Chern insulator case, when $N$ is small, a defect couples the two counter-propagating edge states, reducing the efficiency.

\begin{figure}
\centering
\includegraphics[width=0.5\textwidth]{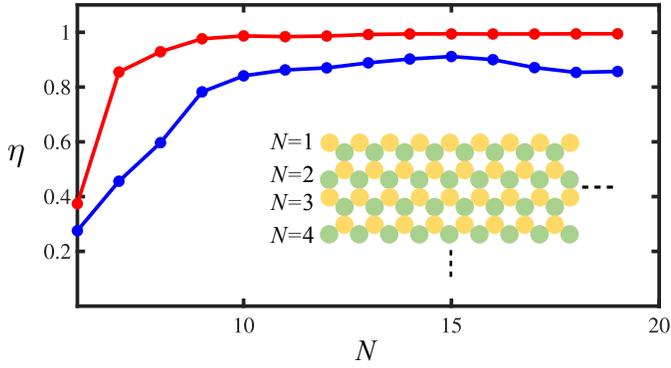}
\caption{Transmission efficiency comparison between anti-chiral edge states (in blue) and the Chern insulator edge states (in red) with different number of layers $N$, as defined in the inset. The Chern insulator can reach almost perfect transmission without backscattering when number of layers is larger than 10. Understandably, the efficiency of antichiral edge states remains lower than that of the Chern insulator, but still reaches around $90\%$. Both types of system has a site missing at the edge.}
\label{Fig6}
\end{figure}

\section{Experimental proposal}
\begin{figure*}[t]
\includegraphics[width=\textwidth]{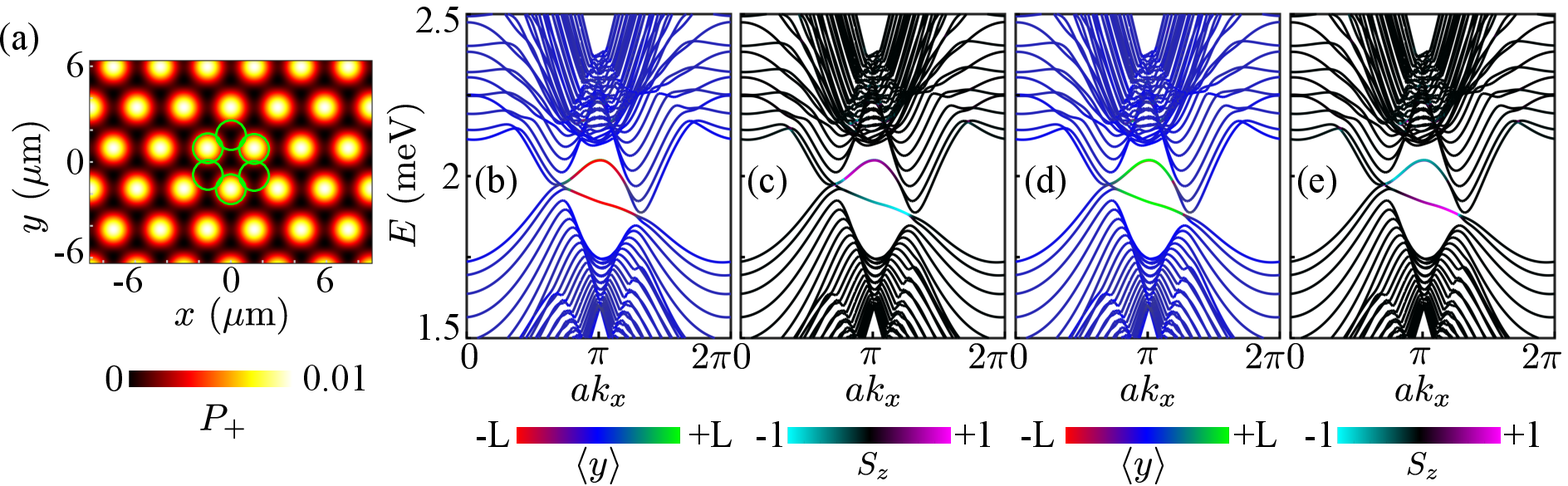}
\caption{(a) Spatial profile of $P_+$ with peaks occurring in only one type of sublattices. The green circles represent the micropillars. (b-e) Band structure of the system. All four of them corresponds to the same system. For presentation purpose, upper edge states are omitted in (b-c) and lower edge states are omitted in (d-e). The bulk is linearly polarized, while the opposite edge states have opposite spins.}
\label{Fig7}
\end{figure*}    
In this section, we discuss and propose a scheme to realize the spin-polarized antichiral edge states in the experiments. The key elements in our scheme are 1) honeycomb lattice potential for the polaritons, 2) TE-TM splitting, and 3) sublattice dependent Zeeman splitting. Arranging exciton-polariton micropillars in a variety of periodic lattices is a routine task in experiments. The TE-TM splitting is known to occur naturally in microcavities due to the polarization dependent reflection from the cavity mirrors \cite{PhysRevB.59.5082}. A Chern insulator was realized based on these ingredients, when a honeycomb lattice was subjected to a strong perpendicular magnetic field, which induced a uniform Zeeman splitting \cite{Klembt2018}. In order to realize the antichiral system, instead of a uniform Zeeman splitting, the sign of the Zeeman splitting needs to be opposite for different sublattices. For this reason, the set up with real magnetic field used for the realization of Chern insulators seems challenging and requires an alternative way. In Ref.~\cite{PhysRevApplied.12.064028}, an additional layer of ferromagnetic material was proposed to induced the Zeeman splitting. Instead of one, two ferromagnetic material layers with opposite predefined magnetic momentums may be used and etched to make the Zeeman splitting opposite in different sublattices. 

However, using the inherent nonlinearity of the polaritons seems to be the most straight-forward way. In a honeycomb lattice under the nonpolarized incoherent pump, it is possible to form polariton condensate in an antiferromagnetic configuration, where different types of sublattices attain opposite circular polarization \cite{PhysRevB.100.235444}. In such a scenario, due to the nonlinear interactions, polaritons with opposite spins will experience blueshifts in different sublattices, giving rise to the necessary sublattice dependent Zeeman splitting in our scheme. 

Another approach could be using circularly polarized incoherent pumps. It was recently verified experimentally that in micropillar structures, a circularly polarized incoherent pump can give rise to optically induced Zeeman splitting up to 0.2 meV \cite{PhysRevResearch.3.043161}. In what follows, we implement this recent experimental finding in our scheme and realize the spin-polarized antichiral edge states. Polaritons under the effect of external incoherent pump can be described using the following Gross-Pitaevskii equation (GPE) \cite{RevModPhys.85.299}

\begin{align}\label{Eq7}
i\hbar\frac{\partial\psi_\pm}{\partial t}=&\left[-\frac{\hbar^2\nabla^2}{2m_o}+V(x,y)+gP_{\pm}(x,y)\right]\psi_\pm\notag\\
&+\left(\alpha_1|\psi_\pm|^2+\alpha_2|\psi_\mp|^2-i\alpha_{NL}|\psi_\pm|^2\right)\psi_\pm\notag\\
&+i\left[P_\pm(x,y)-\Gamma\right]\psi_\pm+\Delta_T\left(i\frac{\partial}{\partial x}\pm\frac{\partial}{\partial y}\right)^2\psi_\mp.
\end{align}
Here $\psi_\pm$ is the polariton wave function corresponding to the ``$\pm$" spins, $m_o$ is the polartion effective mass, and $V(x,y)$ is the underlying honeycomb lattice potential. $gP_{\pm}$ is the spin dependent potential experienced by the polaritons due to interaction with the excitonic reservoir created by the circularly polarized incoherent pumps $P_{\pm}$ with $g$ being a dimensionless parameter. $\Delta_T$ is the TE-TM splitting. $\alpha_{1(2)}$ is the polariton-polariton interaction coefficient between same (opposite) spins. $\alpha_{NL}$ is the nonlinear loss and $\Gamma$ is the linear decay. 

The incoherent pump can be patterned into a lattice shape. One way is to use a spatial light-modulator, which has been used to make lattice potentials in a variety of works~\cite{PhysRevLett.119.067401, PhysRevB.97.195109,Topfer:21,PhysRevApplied.18.024028}. An alternative technique is to interference plane waves with different in-plane wavevector components which has been demonstrated experimentally in \cite{PhysRevApplied.13.044052}. Here in this work, we use superposition of four waves to prepare the incoherent pumps, such that the resultant wave has maxima at one type of sublattice, while minima at the other type of sublattices. Here we emphasize that the four waves we are using here come from the same laser by using additional external optics devices. In this case, $P_+$ can be  expressed as
\begin{align}
P_+(x,y)=P_0f(x,y),
\end{align}
where 
\begin{align}
f(x,y) = &\cos\left(K_1 \cdot R\right)+\cos\left(K_2 \cdot R\right)+\cos\left(K_3 \cdot R\right)\notag\\&+\cos\left(K_4 \cdot R\right)
\end{align}
with $K_1 = 4\pi(0,1/\sqrt3)/a$, $K_{2,3} = 2\pi(1,\pm1/\sqrt3)/a$, $K_4 = (0,0)$.
In the above $a$ is the periodicity of the lattice along the $x$ direction, $P_0$ controls the amplitude of the pump, $f$ is the superposition of four waves that creates the maxima at a particular type of sublattices, and the last term is a wave which has zero in-plane wave vector that fixes the minima of $P_+$ to zeros. In Fig.~\ref{Fig7}(a) the spatial profile of $P_+$ is plotted, which shows that indeed  the maxima of $P_+$ coincide with only one type of sublattices. $P_-$ also has a similar spatial form, however, one needs to shift $f$ spatially such that the maxima now coincide with the other type of sublattice. 

We choose the pillar diameter to be $2~\mu$m, the lattice periodicity around $a=2.95~\mu$m~\cite{PhysRevLett.120.097401, Su:2020wu,PhysRevLett.112.116402}, effective mass $m_0=5\times 10^{-5} m_e$, where $m_e$ is the free electron mass, and the potential depth of the micropillars to be around 4 meV with 0 meV being the minimum. We also fix $\Delta_T=0.12$ meV.$\mu$m$^2$~\cite{Klembt2018,PhysRevLett.115.246401,Sedov:2019wb} and $\Gamma=5.9~\mu$eV~\cite{PhysRevX.3.041015}, which corresponds to a polariton lifetime around 55 ps. By fixing $P_0=2.3~\mu$eV we stay near the condensation threshold, such that $|\psi_\pm|^2\approx$0. This allows us to be in the linear limit and calculate the band structure of the system using the Bloch theorem. We also set $g=2.3$, which induces optical Zeeman shift of the lower energy modes around 0.2 meV, which is consistent with experiment \cite{PhysRevResearch.3.043161}. 

Now that we have all the ingredients, by applying the Bloch theorem to the linear Hamiltonian corresponding to the Eq.~(\ref{Eq7}), we calculate the band structure of the system corresponding to a strip geometry with zigzag edges as shown in Figs.~\ref{Fig7}(b-e). Similar to the tight binding model, our continuous model also predicts the antichiral band structure, where the Dirac points shifts in energy and two pairs of edge states appear. As already discussed, the opposite edge states in our system are degenerate. Consequently, to clearly show that opposite edge states have  opposite spins, we plot the same band structure four times. In Figs.~\ref{Fig7}(b-c) we have omitted the edge states located at the upper edge of the strip ($\langle y\rangle=+L$), while  in Figs.~\ref{Fig7}(d-e) we have omitted the edge states located at the lower edge of the strip ($\langle y\rangle=-L$). Indeed, the opposite edge states attain opposite spins, while the bulk remains linearly polarized. Although, the topological properties of the continuous model band structure is the same with those from the tight-binding band structure, there are a few differences. For example, the upper of pair edge states does not have similar dispersion as the lower pair.  This may arise due to the next nearest neighbour hopping, which is naturally present in the continuous model as well as due to the continuous nature of the effective Zeeman and TE-TM splittings.

The parameters that we have used are flexible. However, the TE-TM splitting and effective sublattice dependent Zeeman splitting are the crucial parameters. For the present parameters, the energy window, where the antichiral edge states resides is around 0.12 meV. This is similar to the topological bandgap obtained experimentally for the polariton Chern insulator \cite{Klembt2018}. Being topological in nature, a slight change in the parameters will not hamper the antichiral modes. However, the case where the Zeeman splitting decreases will  result in a decrease in the energy window of interest. 

For simplicity, we have only considered only the necessary terms in the GPE in Eq.~(\ref{Eq7}) to obtain the desired effect. For example, we have ignored the spin relaxation of the excitonic reservoir and assumed it to preserve their circular polarization perfectly. Although this is not entirely true, experiment has shown that under circularly polarized incoherent pump the reservoir can have up to 17\% preferred $S_z$ \cite{CarlonZambon2019}, which should be enough as long as the blueshift difference  between opposite circularly polarized polariton modes in a pillar is around 0.2 meV.

\section{Discussion and conclusion}
The spin degree of freedom of exciton-polaritons makes them suitable for optical spintronic devices. Devices based on polariton spins have been realized experimentally, which includes spin switches \cite{Amo2010}, logic gates \cite{doi:10.1063/1.4926418}, etc. However, for the realization of a complete optical network, it is important to be able to communicate among different elements without losing the spin information. Unfortunately, the Chiral polaritonic edge modes do not preserve spins while propagating. Our proposed antichiral edge states could be useful in this purpose. Besides all the edge modes in our system propagate in one direction, which may be helpful in transferring oneway information through both the edges, unlike in the Chern insulator where only one edge can be used.  

We note that antichiral edge states for exciton-polaritons were proposed in a previous work \cite{PhysRevB.99.115423}. The present work is significantly different from the previous one in many ways. 
First, in the previous work the antichiral edge modes were not coherent polaritonic states. Instead, those were fluctuations on top of a steady state. Second, in Ref. \cite{PhysRevB.99.115423} the antichiral edge states located at opposite edges did not have opposite spins. Last, in the previous work, the antichiral edge states were shown to propagate only in a straight strip, while their propagation around different types of bendings was not discussed.

To conclude, we have presented a theoretical scheme to obtain spin-polarized antichiral edge states for the first time. We use the naturally present TE-TM splitting and sublattice depend Zeeman splitting in a honeycomb lattice of exciton-polariton micropillar, where the Dirac points shift in energy and two pairs of antichiral edge states appear that propagate in the same directions. While the bulk of the system is linearly polarized, states located at the opposite edges have opposite spins. Although the antichiral edge states reside together with the bulk modes, we found that they show fair robustness and can go around 60 degree bend without being reflected. Our work may be useful in transferring spins and connecting different spin dependent polariton logic elements.

\section{Acknowledgements}
The work was supported by the Ministry of Education, Singapore (Grant No. MOE2019-T2-1-004).

\section{Appendix}
Here in the appendix we discuss how to calculate the winding number in detail.  Different from the model considered in \cite{PhysRevLett.120.086603}, which has only one band below the Dirac points, our present model involves two bands. This makes the Hamiltonian complex and restricts us from calculating the winding number analytically. Additionally, because of the presence of two bands the Berry connections becomes non-Abelian (a matrix instead of a scalar) and one must rely on numerical techniques to calculate the winding number. 

To make the integral Gauge invariance, we use the Wilson loop approach \cite{doi:10.1126/science.aah6442}. We first transform the hexagonal BZ to a rhombic shape with its axes aligned with the $k_1$ and $k_2$ direction as shown in Fig.~\ref{A1}.We notice that such a choice of the BZ is not unique, and the directions $(k_1,k_2)$ used in \cite{PhysRevLett.120.086603} is related to ours by a rotation of $\pi/12$. However, we choose such a BZ as the Dirac points are situated at the central of the BZ when projected on any of the two directions. The Wilson loop operator along $k_2$ for a fixed value of $k_1$ can be defined as:

\begin{figure}
\includegraphics[width=0.3\textwidth]{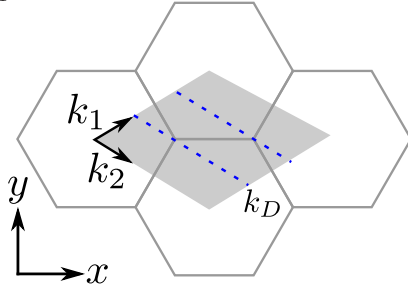}
\caption{A rhombic shape Brillouin zone with its axis aligned with the $k_1$ and $k_2$ direction.}
\label{A1}
\end{figure}

\begin{align}
W(k_1) = &F(k_1,k_2+(N-1)\Delta k)F(k_1,k_2+(N-2)\Delta k)...\notag\\ &F(k_1,k_2+\Delta k)F(k_1,k_2),
\end{align}
where $F$ is a matrix whose elements are given by
\begin{equation}
\left[F(k_1,k_2)\right]^{mn} = \left\langle u^m(k_1,k_2 + \Delta k)|u^n(k_1,k_2)\right\rangle.
\end{equation}
Here $\Delta k$ is the grid spacing in the reciprocal space, the integral is over the unit cell, $u$ is the Bloch state and $m,n = 1,2$ corresponds to the lowest two bands. Next we define a Wannier Hamiltonian $H_{W(k_1)}$ as 
\begin{equation}
W(k_1) = e^{iH_{W(k_1)}}.
\end{equation}
The eigenvalues of $H_{W(k_1)}$ gives the winding numbers for each $k_1$ as shown in Fig. \ref{Fig2}(h) in the main text.

\bibliography{main}

\end{document}